\documentclass [11pt,onecolumn]{mn2e}

\title[Eccentricity generation in HTS]{Eccentricity generation in
hierarchical triple systems: the planetary regime}
\author[Nikolaos Georgakarakos]{Nikolaos Georgakarakos\\100 Delphon Str., Thessaloniki 546 43, Greece\\
email: georgakarakos@hotmail.com}
\date{}

\usepackage{graphicx}

\begin{document}
\maketitle

\begin{abstract}

In previous papers, we developed a technique for estimating the inner
eccentricity in hierarchical triple systems, with the inner
orbit being initially circular.  We considered systems with well 
separated components and different initial setups (e.g. coplanar 
and non-coplanar orbits). However, the systems we examined had  
comparable masses.  In the present paper, the validity of some of 
the formulae derived previously is tested by numerically integrating 
the full equations of motion for systems with smaller mass ratios (from 
${10^{-3} \hspace{0.2cm} \mbox{to} \hspace{0.2cm} 10^{3}}$, i.e. systems with Jupiter-sized bodies).
There is also discussion about HD217107 and its planetary companions.
\end{abstract}


\section{INTRODUCTION}

A hierarchical triple system consists of a binary system and a third body on a
wider orbit.  The motion of such a system can be pictured as the motion
of two binaries on slowly perturbed Keplerian orbits: the binary itself 
(inner binary) and the binary which
consists of the third body and the centre of mass of the binary (outer binary).  
Hierarchical triple systems are widely present in the
galactic field and in star clusters and studying the dynamical evolution of such
systems is a key to understanding a number of issues in astronomy and
astrophysics, such as tidal friction
and dissipation, mass transfer and  mass loss due to a stellar wind,
which may result in changes in stellar
structure and evolution (for systems with {\sl close} inner
binaries, where the separation between the components is
comparable to the radii of the bodies).  But even in
systems with well-separated inner binary components, the perturbation
of the third body can have a devastating effect on the 
triple system as a whole (e.g. disruption of the system).     

For most hierarchical triple stars, the period ratio
${X}$ is of the order of 100 and these systems are probably very stable
dynamically.  However, there are systems with much smaller period
ratios, like the system HD 109648 with ${X=22}$ (Jha et al. 2000), the
${\lambda}$ Tau system, with
\begin{math}
X=8.3
\end{math}
(Fekel ${\&}$ Tomkin 1982)
and the CH Cyg system with
\begin{math}
X=7.0
\end{math}
(Hinkle et al. 1993). 

In two previous papers (Georgakarakos 2002,2003, hereafter HTS1 and 
HTS2 respectively), we derived formulae for the
inner eccentricity in hierarchical triple systems with coplanar and
initially circular inner orbit.  However, the formulae were only tested 
for systems with comparable masses within the range ${10:1}$ (stellar systems). 
In the present paper, the testing is extended to systems with masses within
the range ${1000:1}$, i.e. systems with Jupiter-sized bodies.

\section{THE ECCENTRICITY FORMULAE }
\label{tef}

At this point, we would like to remind the reader of the formulae that were 
derived in HTS1 and HTS2.
The formula for the circular outer binary case is (with the 
addition of two more short period terms, i.e. the next order 
terms in ${P_{2}}$ and ${P_{3}}$ Legendre 
polynomials; those terms which were included in HTS2 for greater accuracy 
and were denoted as ${P_{22}}$ and ${P_{32}}$, can be obtained by setting 
${e=0}$ in equations (9), (11), (13) and (15) of HTS2):
\begin{equation}
\overline{e_{1}^{2}}=\frac{m_{3}^{2}}
{M^{2}}\frac{1}{X^{4}}\left(\frac{43}{4}+
\frac{225}{128}m^{2}_{*}X^{\frac{2}{3}}+\frac{365}{9}\frac{1}{X^{2}}+
\frac{8361}{2048}m^{2}_{*}\frac{1}{X^{\frac{4}{3}}}
+\frac{122}{3}\frac{1}{X}\right)
+\frac{15}{8}\frac{m_{3}}{M}\frac{m_{*}}{X^{\frac{5}{3}}}\frac{C}{A-B}+2\left(\frac{C}{A-B}\right)^{2}\label{final1}
\end{equation}
with
\begin{displaymath}
A=\beta,\hspace{0.3cm}
B=1+\frac{75}{8}\gamma,\hspace{0.3cm}
C=\frac{5}{4}\alpha e_{{\rm T}}, \hspace{0.3cm} e_{{\rm T}}=\frac{3}{4}\frac{m_{1}m_{2}}{(m_{1}+m_{2})^{\frac{4}{3}}M^{\frac{2}{3}}}\frac{1}{X^{\frac{4}{3}}}.
\end{displaymath}

The formula for the eccentric outer binary case is (slightly different from the one in HTS2, as some coefficients have been
corrected, see Georgakarakos 2005):

\begin{eqnarray}
\overline{e_{1}^{2}} & = &
\frac{m_{3}^{2}}{M^{2}}\frac{1}{X^{4}(1-e^{2})^{\frac{9}{2}}}\left[\frac{43}{8}+\frac{129}{8}e^{2}+\frac{129}{64}e^{4}+\frac{1}{(1-e^{2})^{\frac{3}{2}}}(\frac{43}{8}+\frac{645}{16}e^{2}+\frac{1935}{64}e^{4}+\frac{215}{128}e^{6})+\frac{1}{X^{2}(1-e^{2})^{3}}[\frac{365}{18}+\right.\nonumber\\
& &
+\frac{44327}{144}e^{2}+\frac{119435}{192}e^{4}+\frac{256105}{1152}e^{6}+\frac{68335}{9216}e^{8}+\frac{1}{(1-e^{2})^{\frac{3}{2}}}(\frac{365}{18}+\frac{7683}{16}e^{2}+\frac{28231}{16}e^{4}+\frac{295715}{192}e^{6}+\frac{2415}{8}e^{8}+\nonumber\\
& &
+\frac{12901}{2048}e^{10})]+\frac{1}{X(1-e^{2})^{\frac{3}{2}}}[\frac{61}{3}+\frac{305}{2}e^{2}+\frac{915}{8}e^{4}+\frac{305}{48}e^{6}+\frac{1}{(1-e^{2})^{\frac{3}{2}}}(\frac{61}{3}+\frac{854}{3}e^{2}+\frac{2135}{4}e^{4}+\frac{2135}{12}e^{6}+\nonumber\\
& &
+\frac{2135}{384}e^{8})]+m_{*}^{2}X^{\frac{2}{3}}(1-e^{2})[\frac{225}{256}+\frac{3375}{1024}e^{2}+\frac{7625}{2048}e^{4}+\frac{29225}{8192}e^{6}+\frac{48425}{16384}e^{8}+\frac{825}{2048}e^{10}+\frac{1}{(1-e^{2})^{\frac{3}{2}}}(\frac{225}{256}+\nonumber\\
& &
+\frac{2925}{1024}e^{2}+\frac{775}{256}e^{4}+\frac{2225}{8192}e^{6}+\frac{25}{512}e^{8})]+m_{*}^{2}\frac{1}{X^{\frac{4}{3}}(1-e^{2})^{2}}[\frac{8361}{4096}+\frac{125415}{8192}e^{2}+\frac{376245}{32768}e^{4}+\frac{41805}{65536}e^{6}+\nonumber\\
& & \left.+\frac{1}{(1-e^{2})^{\frac{3}{2}}}(\frac{8361}{4096}+\frac{58527}{2048}e^{2}+\frac{877905}{16384}e^{4}+\frac{292635}{16384}e^{6}+\frac{292635}{524288}e^{8})]\right]+2(\frac{C}{B-A})^{2}.\label{final2}
\end{eqnarray}
with
\begin{displaymath}
A=\frac{\beta}{(1-e^{2})^{2}},\hspace{0.3cm}
B=\frac{1}{(1-e^{2})^{\frac{3}{2}}}+\frac{25}{8}\gamma\frac{3+2e^{2}}{(1-
e^{2})^{3}}\hspace{0.2cm} \mbox{and} \hspace{0.1cm}C=\frac{5}{4}\alpha\frac{e}{(1-e^{2})^{\frac{5}{2}}}.
\end{displaymath}

In both cases:
\begin{displaymath}
\alpha =\frac{m_{1}-m_{2}}{m_{1}+m_{2}}\frac{a_{1}}{a_{{2}}},\hspace{0.2cm}\beta
=\frac{m_{1}m_{2}M^{\frac{1}{2}}}{m_{3}(m_{1}+m_{2})^{\frac{3}{2}}}(\frac{a_{{1}}}{a_{{2}}})^{\frac{1}{2}},\hspace{0.2cm}\gamma=\frac{m_{3}}{M^{\frac{1}{2}}(m_{1}+m_{2})^{\frac{1}{2}}}(\frac{a_{{1}}}{a_{{2}}})^{\frac{3}{2}},\hspace{0.5 cm}\\
\end{displaymath}
${e \hspace{0.2cm}\mbox{is the outer eccentricity and} \hspace{0.2cm}a_{1}\hspace{0.2cm}\mbox{and} \hspace{0.2cm} a_{2} \hspace{0.2cm}\mbox{are the inner and outer semi major axes respectively.}}$ 
Also, ${M}$ is the total mass of the system, ${
m_{*}=\frac{m_{2}-m_{1}}{(m_{1}+m_{2})^{\frac{2}{3}}M^{\frac{1}{3}}}}$ 
and ${X}$ is the period ratio of the two orbits.

\section{NUMERICAL TESTING}
In order to test the validity of the formulae derived in the previous
papers, we integrated the full equations of motion numerically, using
a symplectic integrator with time transformation (Mikkola 1997).

The code calculates the relative position and velocity vectors of the
two binaries at every time step.  Then, by using standard two body formulae,
we computed the orbital elements of the two binaries.
The various parameters used by the code, were given the following
values: writing index ${Iwr=1}$, method coefficients ${a1=1}$ and ${a2=15}$, correction index ${icor=1}$.  
The average number of steps per inner binary period ${NS}$, was given the value of ${60}$ for testing the 
short period terms and the long term circular case. For the long period eccentric cases, we set ${NS=5}$ in order to accelerate
 the execution of the code, without any precision cost, as the motion was mainly dominated by 
secular evolution.

For our simulations, we also defined the two mass ratios 
\begin{displaymath}
K1=\frac{m_{1}}{m_{1}+m_{2}} \hspace{0.2cm}\mbox{and} \hspace{0.2cm}
K2=\frac{m_{3}}{m_{1}+m_{2}}, 
\end{displaymath}
with ${0.001 \leq K1 \leq 0.5}$ and ${0.001 \leq K2 \leq 1000}$.
 
Following HTS2, for the eccentric outer binary case, we used the 
fictitious initial period ratio ${X_{0f}}$, defined as
the ratio of the period that the outer binary would have on a circular
orbit with a semi major axis equal to its periastron distance over the
period of the inner binary.  In all cases ${X_{0f} \geq 10}$.  
We also used units such that
${G=1}$ and ${m_{1}+m_{2}=1}$ and we always started the integrations
with ${a_{1}=1}$.  In that system of units, the initial conditions for 
the numerical integrations were as follows:

\begin{displaymath}
r_{1}=1,\hspace{0.5cm} r_{2}=0,\hspace{0.5cm} r_{3}=0
\end{displaymath}	
\begin{displaymath}
R_{1}=a_{2}\cos{\phi},\hspace{0.5cm} R_{2}=a_{2}\sin{\phi},\hspace{0.5cm} R_{3}=0
\end{displaymath}	
\begin{displaymath}
\dot{r}_{1}=0,\hspace{0.5cm} \dot{r}_{2}=1,\hspace{0.5cm} \dot{r}_{3}=0
\end{displaymath}	
\begin{displaymath}
\dot{R}_{1}=-\sqrt{\frac{M}{a_{2}}}\sin{\phi},\hspace{0.5cm} \dot{R}_{2}=\sqrt{\frac{M}{a_{2}}}\cos{\phi},\hspace{0.5cm} \dot{R}_{3}=0,
\end{displaymath}

for the circular case (${\phi}$ is the initial relative phase of the two 
binaries) and

\begin{displaymath}
r_{1}=1,\hspace{0.5cm} r_{2}=0,\hspace{0.5cm} r_{3}=0
\end{displaymath}	
\begin{displaymath}
R_{1}=R_{0}\cos{(f_{0}+\varpi)},\hspace{0.5cm} R_{2}=R_{0}\sin{(f_{0}+\varpi)},\hspace{0.5cm} R_{3}=0
\end{displaymath}	
\begin{displaymath}
\dot{r}_{1}=0,\hspace{0.5cm} \dot{r}_{2}=1,\hspace{0.5cm} \dot{r}_{3}=0
\end{displaymath}	
\begin{displaymath}
\dot{R}_{1}=-\sqrt{\frac{M}{a_{2}(1-e^{2})}}\sin{(f_{0}+\varpi)},\hspace{0.5cm} \dot{R}_{2}=\sqrt{\frac{M}{a_{2}(1-e^{2})}}\cos{(f_{0}+\varpi)},\hspace{0.5cm} \dot{R}_{3}=0.
\end{displaymath}

for the non circular case, where ${f_{0}}$ and ${\varpi}$ are the initial true anomaly and longitude of pericentre respectively of the outer orbit.

\subsection{LONG PERIOD EVOLUTION}

First, we present the results from testing equations (\ref{final1}) 
and (\ref{final2}) for long term behaviour.  The formulae were 
compared with results obtained 
from integrating the full equations of motion numerically.

For the circular case, each system was numerically integrated for
${\phi=0^{\circ}-360^{\circ}}$ with a step of ${45^{\circ}}$.  After
each run, ${e^{2}_{{\rm in}}}$ was averaged over time using the trapezium
rule and after the integrations for all ${\phi}$ were done, we
averaged over ${\phi}$ by using the rectangle rule.  The integrations
were also done for smaller steps in ${\phi}$ (e.g. ${1^{\circ}}$), 
but was not any difference in the outcome.

For the non circular case, each system was numerically integrated for
${\varpi=0^{\circ}-360^{\circ}}$ and ${f_{0}=0^{\circ}-360^{\circ}}$
with a step of ${60^{\circ}}$.  For a given value of ${\varpi}$ and
${f_{0}}$ we integrated our system. After
each run, ${e^{2}_{{\rm in}}}$ was averaged over time using the trapezium
rule and then we integrated the system for a different ${f_{0}}$.  After the integrations for all ${f_{0}}$ were done, we
averaged over ${f_{0}}$ by using the rectangle rule. Then, the same
procedure was applied for the next value of ${\varpi}$ and when the integrations
for all ${\varpi}$ were done, we averaged over ${\varpi}$ by using the
rectangle rule.   The integrations
were also done for smaller steps in ${\varpi}$ and ${f_{0}}$, but 
the difference in the outcome was insignificant.

These results are presented in Tables 1, 2 and 3 (${e=0, 0.4, 0.75}$
respectively), which give the percentage error between the averaged 
numerical ${e^{2}_{{\rm in}}}$ and equations (\ref{final1}) and 
(\ref{final2}).  The error is accompanied by the
period of the oscillation of the eccentricity, which is the same as
the integration time span (when there is no period given, we integrated
for an outer orbital period, i.e ${2\pi X}$, as there was not any noticeable 
secular evolution).  There are four values per
(${K1-K2}$) pair, corresponding, from top to bottom, to ${X=10,20,30,50}$
respectively (of course, for the eccentric outer binary case, ${X}$ is
replaced by ${X_{0f}}$). 

Generally, it appears that the theory is in agreement with the
numerical integrations.  There are some cases (larger ${K2}$) where the 
error is rather
significant (around ${30 \%}$; however, the reader should have in mind 
that the error is in ${e^{2}}$, which means that it is about double than 
the one in ${e}$), but it dropped as the period ratio increased.  There are
also some cases where the formulae do not seem to work well; in fact, for some
systems (e.g. ${e=0.75}$, ${K1=0.005}$,${K2=0.001}$)  we have a complete
failure of our formula.  For those cases, our formula gives an overestimate
of the inner eccentricity (our secular solution has larger period and 
amplitude of oscillation) and that is due to the fact that ${A-B}$ gets small,
 i.e. our approximate secular solution is near resonance.
This is demonstrated in figures 1,2 and 3, which are plots of ${K2}$ 
against ${K1}$ for which ${A-B=0}$.  Finally, for large eccentricities, there
is a significant error for ${K1=0.5}$ and ${X_{0f}=10}$, which is due to terms
omitted from the approximate secular solution.  Those terms become insignificant 
as ${X_{0f}}$ increases.

\begin{table}
\caption[]{Percentage error between the
averaged numerical ${e^{2}_{1}}$ and equation (\ref{final1}).  
For all systems, ${e=0}$.}
\vspace{0.1 cm}
\begin{center}	
{\normalsize \begin{tabular}{c c c c c c c }\hline
${K2\backslash\ K1}$ & ${0.001}$ & ${0.005}$ & ${0.01}$ & ${0.05}$ & ${0.1}$ & ${0.5}$ \\
\hline
0.001 & -4.5 (${1.15 10^{6}}$)& -3 (${4.7 10^{5}}$)     & 6.9 (${2 10^{5}}$)  & 9.4 (${3.5 10^{4}}$)& 9.7 (${1.9 10^{4}}$) & 12.2 \\
      & -1.5 (${5 10^{6}}$)   & -15.7 (${3.5 10^{6}}$)  & 2.3 (${1.1 10^{6}}$)& 3  (${1.9 10^{5}}$) & 3.1 (${ 10^{5}}$)& 3.7 \\
      & 0.4 (${1.05 10^{7}}$) & -16.8 (${1.05 10^{7}}$) & 2.4 (${3 10^{6}}$)  & 1.7 (${5 10^{5}}$)  & 1.7 (${2.6 10^{5}}$)& 1.9 \\
      & -1.2 (${3 10^{7}}$)   & -29.4 (${5 10^{7}}$)    & -0.4 (${12 10^{6}}$)& 0.9 (${1.7 10^{6}}$)& 0.9 (${8.5 10^{5}}$)& 0.8 \\
      
0.01 & 10.2 & 8.3 (${10^{5}}$)     & -4  (${1.2 10^{5}}$) & -1.7 (${5.3 10^{4}}$) & 8.3(${2.3 10^{4}}$) & 12.2 \\
     & 2.1  & 3.9 (${3.6 10^{5}}$) & -0.9 (${5 10^{5}}$)  & -15.9 (${4 10^{5}}$) & 2.8 (${1.3 10^{5}}$)& 3.8 \\
     & 0.3  & 2.7 (${8 10^{5}}$)   & -0.5 (${1.1 10^{6}}$)& -17 (${1.2 10^{6}}$) & 1.6 (${3.7 10^{5}}$)& 1.9 \\
     & -0.8 & 0.5 (${2.5 10^{6}}$) & 0.7  (${2.8 10^{6}}$)& -37 (${6.1 10^{6}}$)& -2.2 (${1.5 10^{6}}$)& 0.8 \\
   
0.1  & 13.8 &13.3 & 12.6                & 11.4 (${9 10^{3}}$) & 7 (${1.1 10^{4}}$) & 12.8 \\
     & 5.4  & 4.7 & 5.1 (${3.5 10^{4}}$)& 4.6 (${4 10^{4}}$)  & 2.8 (${5 10^{4}}$) & 4 \\
     & 3.3  & 2.5 & 3 (${9 10^{4}}$)    & 3.1 (${9 10^{4}}$)  & 2.2 (${1.1 10^{5}}$)& 2.1 \\
     & 1.8  & 1   & 1.8 (${2.2 10^{5}}$)& 1.9 (${2.5 10^{5}}$)& 1.2 (${3 10^{5}}$) & 1 \\
    
1    & 26.7 & 26.6 &26.4  & 25   & 23.5                 & 17.3 \\
     & 12.8 & 12.7 & 12.6 & 11.5 & 10.2                 & 5.9 \\
     & 8.8  & 8.7  & 8.6  & 7.6  & 6.8 (${1.5 10^{4}}$) & 3.2 \\
     & 5.6  & 5.5  & 5.4  & 4.5  & 4.3 (${4.5 10^{4}}$) & 1.6 \\

10  & 36    & 35.9  &35.7  & 34.5 &32.9 &26.2  \\
    & 14.7  & 14.7  & 14.6 & 13.8 &12.8 & 8.4  \\
    & 9.8   & 9.7   & 9.6  & 8.9  & 8.2 & 4.6 \\
    & 6.2   & 6.2   & 6.1  & 5.6  & 5.4 & 2.2 \\
  
100  & 31.4 & 31.4 &31.4 & 30.9  & 30.4 & 28.3 \\
     & 10.9 & 10.8 &10.8 & 10.5  & 10.2 & 8.9 \\
     & 6.5  & 6.4  & 6.4 & 6.2   & 5.9  & 4.8 \\
     & 3.7  & 3.7  & 3.6 & 3.5   & 3.2  & 2.3 \\
   
1000 & 29.3 & 29.4 &29.2 & 29.1 & 29 & 28.5 \\
     & 9.4  & 9.4  & 9.4 & 9.3  & 9.2 & 8.9 \\
     & 5.2  & 5.2  & 5.2 & 5.2  & 5.1 & 4.8 \\
     & 2.7  & 2.7  & 2.7 & 2.6  & 2.6 & 2.4 \\

\hline
\end{tabular}}
\end{center}	
\end{table}

\begin{table}
\caption[]{Percentage error between the
averaged numerical ${e^{2}_{1}}$ and equation (\ref{final2}).  
For all systems, ${e=0.4}$.}
\vspace{0.1 cm}
\begin{center}	
{\normalsize \begin{tabular}{c c c c c c c }\hline
${K2\backslash\ K1}$ & ${0.001}$ & ${0.005}$ & ${0.01}$ & ${0.05}$ & ${0.1}$ & ${0.5}$ \\
\hline
0.001 & -8.5 (${4.73 10^{6}}$)& -35.8 (${2.5 10^{6}}$)  & -8.1 (${9.5 10^{5}}$) & -0.9 (${1.61 10^{5}}$)&  0.2 (${8.3 10^{4}}$)  & 12.9 (${1.5 10^{4}}$) \\
      & -2.8 (${1.72 10^{7}}$)& -40.6 (${1.8 10^{7}}$)  & -4.9 (${5.5 10^{6}}$) & -0.6 (${8.51 10^{5}}$)& -0.1 (${4.35 10^{5}}$) & 3.8 (${8 10^{4}}$) \\
      & -2   (${3.7 10^{7}}$) & -55.5 (${6.04 10^{7}}$) & -3.7 (${1.54 10^{7}}$)& -0.3 (${2.23 10^{6}}$)& -0.2 (${1.135 10^{6}}$)& 2 (${2 10^{5}}$) \\
      & -0.9 (${9.7 10^{7}}$) & -171.6 (${3.13 10^{8}}$)& -2.8 (${5.75 10^{7}}$)& -0.2 (${7.5 10^{6}}$) & -0.2 (${3.79 10^{6}}$) & 1.1 (${6.65 10^{5}}$)\\
      
0.01 &  0.4 (${3 10^{5}}$)   & -2.6 (${3.6 10^{5}}$) & -7.8  (${4.7 10^{5}}$)& -36.3 (${2.86 10^{5}}$) & -5.1 (${1.11 10^{5}}$)& 12.9(${1.8 10^{4}}$) \\
     & -0.3 (${1.23 10^{6}}$)& -0.3 (${1.4 10^{6}}$) & -1.7 (${1.71 10^{6}}$)& -42.4 (${2.06 10^{6}}$) & -3.5 (${6.45 10^{5}}$)& 3.8 (${8.5 10^{4}}$)\\
     & -0.1 (${2.77 10^{6}}$)& -0.6 (${3.12 10^{6}}$)& -0.9 (${3.68 10^{6}}$)& -65.5 (${7.07 10^{6}}$) & -2.8 (${1.82 10^{6}}$)& 2.2 (${2.3 10^{5}}$)\\
     & -0.2 (${7.7 10^{6}}$) & -0.4 (${8.5 10^{6}}$) & -0.7 (${9.75 10^{6}}$)& -333.2 (${3.82 10^{7}}$)& -2.7 (${6.94 10^{6}}$)& 1.6 (${7.95 10^{5}}$)\\
   
0.1  & -0.6 (${3 10^{4}}$)  & -1 (${3.05 10^{4}}$)  & -0.4 (${3.08 10^{4}}$)& -2.6 (${3.6 10^{4}}$)& -5.5 (${4.45 10^{4}}$)& 14.6 (${1.9 10^{5}}$)\\
     & 1.6 (${1.26 10^{5}}$)& 1.6 (${1.275 10^{5}}$)& 1.1 (${1.3 10^{5}}$)  & 0.4 (${1.48 10^{5}}$)& -0.4 (${1.75 10^{5}}$)& 6.8  (${2.7 10^{5}}$) \\
     & 1.4 (${2.88 10^{5}}$)& 1 (${2.92 10^{5}}$)   & 1 (${2.96 10^{5}}$)   & 1.1 (${3.3 10^{5}}$) & 0.3 (${3.83 10^{5}}$) & 3.6 (${4.6 10^{5}}$)\\
     & 0.9 (${8.1 10^{5}}$) & 0.6 (${8.2 10^{5}}$)  & 0.5 (${8.3 10^{5}}$)  & 0.6 (${9.1 10^{5}}$) & 0.5 (${1.025 10^{6}}$)& 1.7 (${9.8 10^{5}}$)\\
    
1    & 1.4 (${4.2 10^{3}}$)& 4.2 (${4.05 10^{3}}$)& 5.4 (${4 10^{3}}$)   & 3.5 (${4.1 10^{3}}$)& 4 (${4.1 10^{5}}$)   & 17.7 (${2.4 10^{3}}$)\\
     & 8 (${2 10^{4}}$)    & 7.9 (${2 10^{4}}$)   & 7.1 (${2.02 10^{4}}$)& 9 (${2 10^{4}}$)    & 7.9 (${2.05 10^{4}}$)& 5.6 (${1.1 10^{4}}$)\\
     & 6.3 (${4.8 10^{4}}$)& 6.4 (${4.8 10^{4}}$) & 6.5 (${4.8 10^{4}}$) & 7.5 (${4.8 10^{4}}$)& 6.8 (${4.9 10^{4}}$) & 3.2 (${2.7 10^{4}}$)\\
     & 4.2 (${1.4 10^{5}}$)& 4.3 (${1.4 10^{5}}$) & 4.4 (${1.4 10^{5}}$) & 4 (${1.42 10^{5}}$) & 4.4 (${1.43 10^{5}}$)& 2 (${7.5 10^{4}}$)\\

10  & 16.8 (${3.9 10^{4}}$) & 16.5 (${4 10^{4}}$)  & 16.6 (${4 10^{4}}$)   & 17 (${4 10^{4}}$)     & 17 (${4 10^{4}}$)    & 24.3  (${4.4 10^{3}}$) \\
    & 12.4 (${ 10^{4}}$)    & 14.6 (${9.7 10^{3}}$)& 16.3 (${9.5 10^{3}}$) & 14.7 (${9.7 10^{3}}$) & 13.1 (${9.9 10^{3}}$)& 8  (${5 10^{3}}$) \\
    & 10.9 (${2.45 10^{4}}$)& 12.5 (${2.4 10^{4}}$)& 10.4 (${2.46 10^{4}}$)& 10.9 (${2.45 10^{4}}$)& 12.8 (${2.4 10^{4}}$)& 4.7 (${1.3 10^{4}}$) \\
    & 8.1  (${7.3 10^{4}}$) & 0.2  (${7.3 10^{4}}$)& 8.1 (${7.3 10^{4}}$)  & 8.2 (${7.3 10^{4}}$)  & 8.4 (${7.3 10^{4}}$) & 2.7 (${3.8 10^{4}}$) \\
  
100  & 21.5 (${1.6 10^{3}}$)& 20.7 (${1.6 10^{3}}$)& 20.8 (${1.6 10^{3}}$)& 21.3 (${1.6 10^{3}}$)& 21.9 (${1.6 10^{3}}$)& 27.1 \\
     & 13 (${9 10^{3}}$)    & 12.5 (${9 10^{3}}$)  & 12.5 (${9 10^{3}}$)  & 14.1 (${8.8 10^{3}}$)& 9.3 (${9.4 10^{3}}$) & 8.4 (${5 10^{3}}$) \\
     & 8.5 (${2.3 10^{4}}$) & 12 (${2.2 10^{4}}$)  & 12 (${2.2 10^{4}}$)  & 12 (${2.2 10^{4}}$)  & 12 (${2.2 10^{4}}$)  & 5 (${1.1 10^{4}}$) \\
     & 9.3 (${6.6 10^{4}}$) & 7.8 (${6.7 10^{4}}$) & 7.8 (${6.7 10^{4}}$) & 7.9 (${6.7 10^{4}}$) & 7.8 (${6.7 10^{4}}$) & 2.8 (${3.4 10^{4}}$) \\
   
1000 & 24.5 (${1.6 10^{3}}$)& 24 (${1.6 10^{3}}$)  & 24 (${1.6 10^{3}}$)   & 24.2 (${1.6 10^{3}}$)& 24.5 (${1.6 10^{3}}$) & 27.1 \\
     & 11 (${9 10^{3}}$)    & 13.7 (${8.5 10^{3}}$)& 13.6 (${8.5 10^{3}}$) & 12.7 (${8.6 10^{3}}$)& 10.7 (${8.9 10^{3}}$) & 8.3 (${4.8 10^{3}}$) \\
     & 9 (${2.25 10^{4}}$)  & 13.9 (${2.1 10^{4}}$)& 12.2 (${2.15 10^{4}}$)& 13.6 (${2.1 10^{4}}$)& 10.1 (${2.2 10^{4}}$) & 4.8 (${1.05 10^{4}}$) \\
     & 9.6 (${6.5 10^{4}}$) & 8.2 (${6.6 10^{4}}$) & 8.2 (${6.6 10^{4}}$)  & 8.2 (${6.6 10^{4}}$) & 9.4 (${6.5 10^{4}}$)  & 2.7 (${3.3 10^{4}}$) \\

\hline
\end{tabular}}
\end{center}	
\end{table}

\begin{table}
\caption[]{Percentage error between the
averaged numerical ${e^{2}_{1}}$ and equation (\ref{final2}).  For all systems, ${e=0.75}$.}
\vspace{0.1 cm}
\begin{center}	
{\normalsize \begin{tabular}{c c c c c c c }\hline
${K2\backslash\ K1}$ & ${0.001}$ & ${0.005}$ & ${0.01}$ & ${0.05}$ & ${0.1}$ & ${0.5}$ \\
\hline
0.001 & 6.8 (${2.64 10^{7}}$)& -54.8 (${1.61 10^{7}}$)  & -10.8 (${5.82 10^{6}}$)& -2.6 (${9.5 10^{5}}$) & -0.8 (${4.88 10^{5}}$) & 81.8 (${8.5 10^{4}}$) \\
      & 1.3 (${8.8 10^{7}}$) & -84.9 (${1.16 10^{8}}$)  & -5.9 (${3.4 10^{7}}$)  & -0.5 (${5 10^{6}}$)   & 0.1 (${2.54 10^{6}}$)  & 5.6 (${4.3 10^{5}}$)\\
      & 0.3 (${1.87 10^{8}}$)& -176.2 (${3.91 10^{8}}$) & -4.6  (${9.6 10^{7}}$) & -0.2 (${1.31 10^{7}}$)& -0.2 (${6.64 10^{6}}$) & 2.6 (${1.1 10^{6}}$)\\
      & 0.3 (${4.9 10^{8}}$) & -8233.2 (${1.92 10^{9}}$)& -4.2 (${3.67 10^{8}}$) & -0.3 (${4.42 10^{7}}$)& -0.1 (${2.215 10^{7}}$)& 1.1 (${3.8 10^{6}}$)\\
      
0.01 & 3.6 (${1.63 10^{6}}$)& 4 (${1.93 10^{6}}$)  & 6.6 (${2.6 10^{6}}$) & -55.7 (${1.81 10^{6}}$) & -7.4 (${6.77 10^{5}}$) & 70 (${9 10^{4}}$)\\
     & 0.5 (${6.45 10^{6}}$)& 0.8 (${7.3 10^{6}}$) & 1.4 (${8.8 10^{6}}$) & -100.5 (${1.33 10^{7}}$)& -4.3 (${4 10^{6}}$)    & 5.5 (${5 10^{5}}$)\\
     & 0.7 (${1.44 10^{7}}$)& 0.3 (${1.61 10^{7}}$)& 0.7 (${1.87 10^{7}}$)& -262.9 (${4.55 10^{7}}$)& -3.8 (${1.147 10^{7}}$)& 2.7 (${1.4 10^{6}}$)\\
     & 0.1 (${4 10^{7}}$)   & 0.3 (${4.36 10^{7}}$)& 0.1 (${4.94 10^{7}}$)& -4855 (${2.28 10^{8}}$) & -3.8 (${4.49 10^{7}}$) & 1.2 (${4.6 10^{6}}$)\\
   
0.1  & 1.8  (${1.61 10^{5}}$)& 2.1 (${1.63 10^{5}}$) & 2.1 (${1.66 10^{5}}$) & 1.8 (${1.93 10^{5}}$) & 1.7 (${2.37 10^{5}}$)& 37.8 (${3.3 10^{5}}$)\\
     & 1.3 (${6.65 10^{5}}$) & 0.9 (${6.75 10^{5}}$) & 1.1 (${6.83 10^{5}}$) & 1.3 (${7.65 10^{5}}$) & 0.9 (${8.92 10^{5}}$)& 5.7 (${10^{6}}$)\\
     & 0.8 (${1.515 10^{6}}$)& 1.1 (${1.525 10^{6}}$)& 0.7 (${1.55 10^{6}}$) & 0.9 (${1.71 10^{6}}$) & 0.9 (${1.94 10^{6}}$)& 2.8 (${2 10^{6}}$)\\
     & 0.4 (${4.24 10^{6}}$) & 0.6 (${4.27 10^{6}}$) & 0.5 (${4.32 10^{6}}$) & 0.4 (${4.7 10^{6}}$)  & 0.5 (${5.21 10^{6}}$)& 1.3 (${4.4 10^{6}}$)\\
    
1    & -3.4 (${2.2 10^{4}}$)& -3.3 (${2.2 10^{4}}$)& -3.3 (${2.2 10^{4}}$)& -5.5 (${2.25 10^{4}}$) & -5.5  (${2.25 10^{4}}$)& 36.7 (${1.14 10^{4}}$)\\
     & 2.6 (${1.08 10^{5}}$)& 3.6 (${1.07 10^{5}}$)& 4.6 (${1.06 10^{5}}$)& 2.6 (${1.09 10^{5}}$)  & 2.6  (${1.1 10^{5}}$)  &  7.1 (${5.8 10^{4}}$)\\
     & 3.1 (${2.55 10^{5}}$)& 3.6 (${2.54 10^{5}}$)& 3.7 (${2.54 10^{5}}$)& 3.8 (${2.56 10^{5}}$)  & 3.2  (${2.6 10^{5}}$)  &  3.7 (${1.4 10^{5}}$)\\
     & 2.5 (${7.35 10^{5}}$)& 2.6 (${7.35 10^{5}}$)& 2.3 (${7.38 10^{5}}$)& 2.9 (${7.4 10^{5}}$)   & 2.5 (${7.5 10^{5}}$)   &  2.1 (${3.9 10^{5}}$)\\

10  & 2.7 (${9 10^{3}}$)   & -6.3 (${10^{4}}$)    & -7 (${10^{4}}$)      & -2.6 (${9.5 10^{3}}$)& 3.2 (${9 10^{3}}$)   & 35 (${4.5 10^{3}}$)\\
    & 8.3 (${5.1 10^{4}}$) & 6.6 (${5.2 10^{4}}$) & 6.6 (${5.2 10^{4}}$) & 6.6 (${5.2 10^{4}}$) & 6.6 (${5.2 10^{4}}$) & 9.4 (${2.7 10^{4}}$)\\
    & 7 (${1.28 10^{5}}$)  & 5.6 (${1.3 10^{5}}$) & 5.6 (${1.3 10^{5}}$) & 6.4 (${1.29 10^{5}}$)& 6.5 (${1.29 10^{5}}$)& 5.2 (${6 10^{4}}$)\\
    & 4.8 (${3.85 10^{5}}$)& 4.8 (${3.85 10^{5}}$)& 4.8 (${3.85 10^{5}}$)& 4.2 (${3.88 10^{5}}$)& 4.1 (${3.89 10^{5}}$)& 2.7 (${2 10^{5}}$)\\
  
100  & 11 (${8.5 10^{3}}$)  & 10.8 (${8.5 10^{3}}$)& 10.9 (${8.5 10^{3}}$)& 12.2 (${8.5 10^{3}}$)& 16.4 (${8.5 10^{3}}$)& 31.8 (${4 10^{3}}$)\\
     & 6.3 (${4.7 10^{4}}$) & 4.4 (${4.8 10^{4}}$) & 4.4 (${4.8 10^{4}}$) & 6.3 (${4.7 10^{4}}$) & 2.7 (${4.9 10^{4}}$) & 10 (${2.5 10^{4}}$)\\
     & 6.4 (${1.17 10^{5}}$)& 7.2 (${1.16 10^{5}}$)& 7.2 (${1.16 10^{5}}$)& 6.4 (${1.17 10^{5}}$)& 5.7 (${1.18 10^{5}}$)& 5.8 (${5.8 10^{4}}$) \\
     & 5.3 (${3.5 10^{5}}$) & 5.3 (${3.5 10^{5}}$) & 5.3 (${3.5 10^{5}}$) & 5.3 (${3.5 10^{5}}$) & 5.3 (${3.5 10^{5}}$) & 3 (${1.75 10^{5}}$)\\
   
1000 & 33.5 (${8 10^{3}}$)  & 33.6 (${8 10^{3}}$)  & 33.8 (${8 10^{3}}$)  & 35.3 (${8 10^{3}}$)  & 37 (${8 10^{3}}$)    & 38.8 (${4 10^{3}}$)\\
     & 4.1 (${4.8 10^{4}}$) & 4.1 (${4.8 10^{4}}$) & 4.1 (${4.8 10^{4}}$) & 4.3 (${4.8 10^{4}}$) & 4.6 (${4.8 10^{4}}$) & 10 (${2.5 10^{4}}$)\\
     & 6.2 (${1.16 10^{5}}$)& 7 (${1.15 10^{5}}$)  & 7 (${1.15 10^{5}}$)  & 7 (${1.15 10^{5}}$)  & 7 (${1.15 10^{5}}$)  & 5.7  (${5.8 10^{4}}$)\\
     & 4.9 (${3.48 10^{5}}$)& 4.9 (${3.48 10^{5}}$)& 4.9 (${3.48 10^{5}}$)& 4.9 (${3.48 10^{5}}$)& 4.9 (${3.48 10^{5}}$)& 3 (${1.75 10^{5}}$)\\
\hline
\end{tabular}}
\end{center}	
\end{table}

\begin{figure}
\begin{center}
\includegraphics[width=80mm,height=60mm]{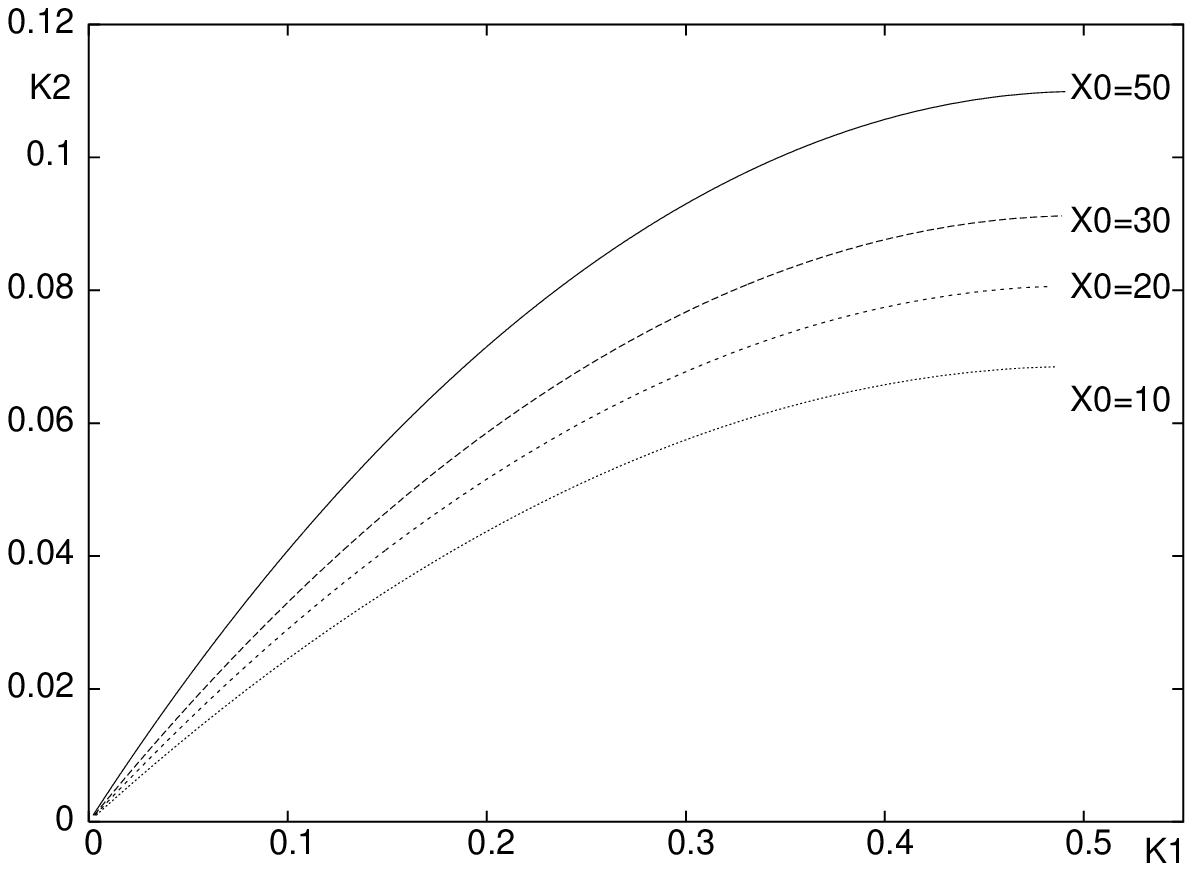}
\includegraphics[width=80mm,height=60mm]{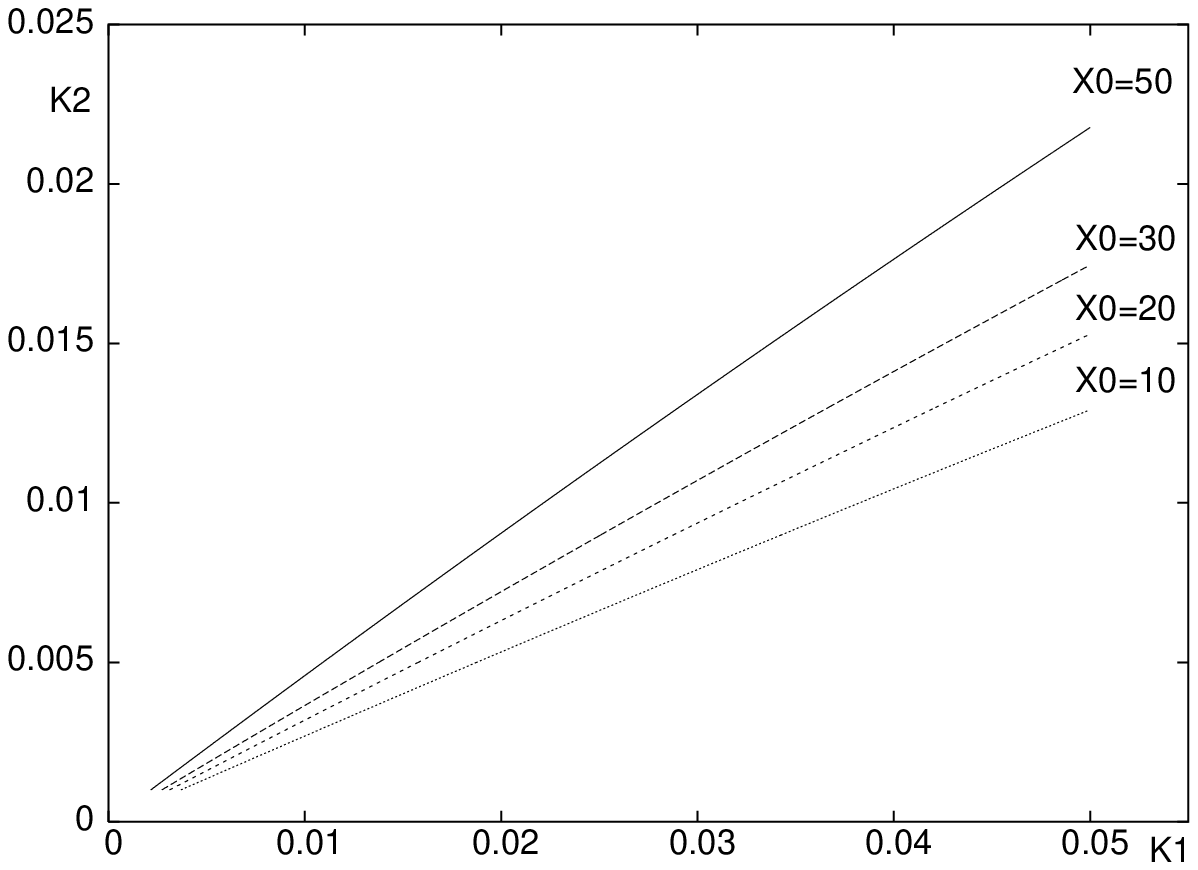}
\caption[]{${K2}$ against ${K1}$ for ${e=0}$ and for which ${A
-B=0}$.  The right graph is the same as the left one, but for 
${K1 \leq 0.05}$.}
\end{center}
\end{figure}

\begin{figure}
\begin{center}
\includegraphics[width=80mm,height=60mm]{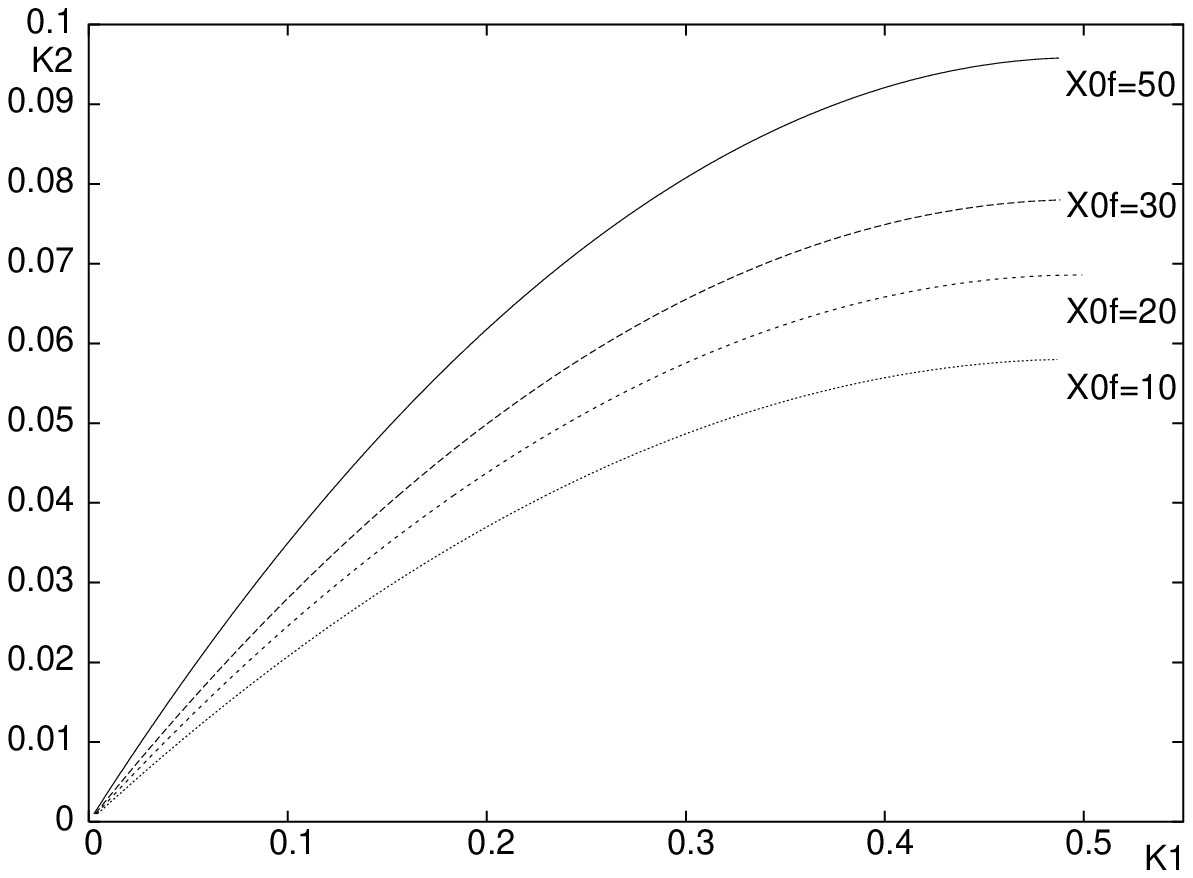}
\includegraphics[width=80mm,height=60mm]{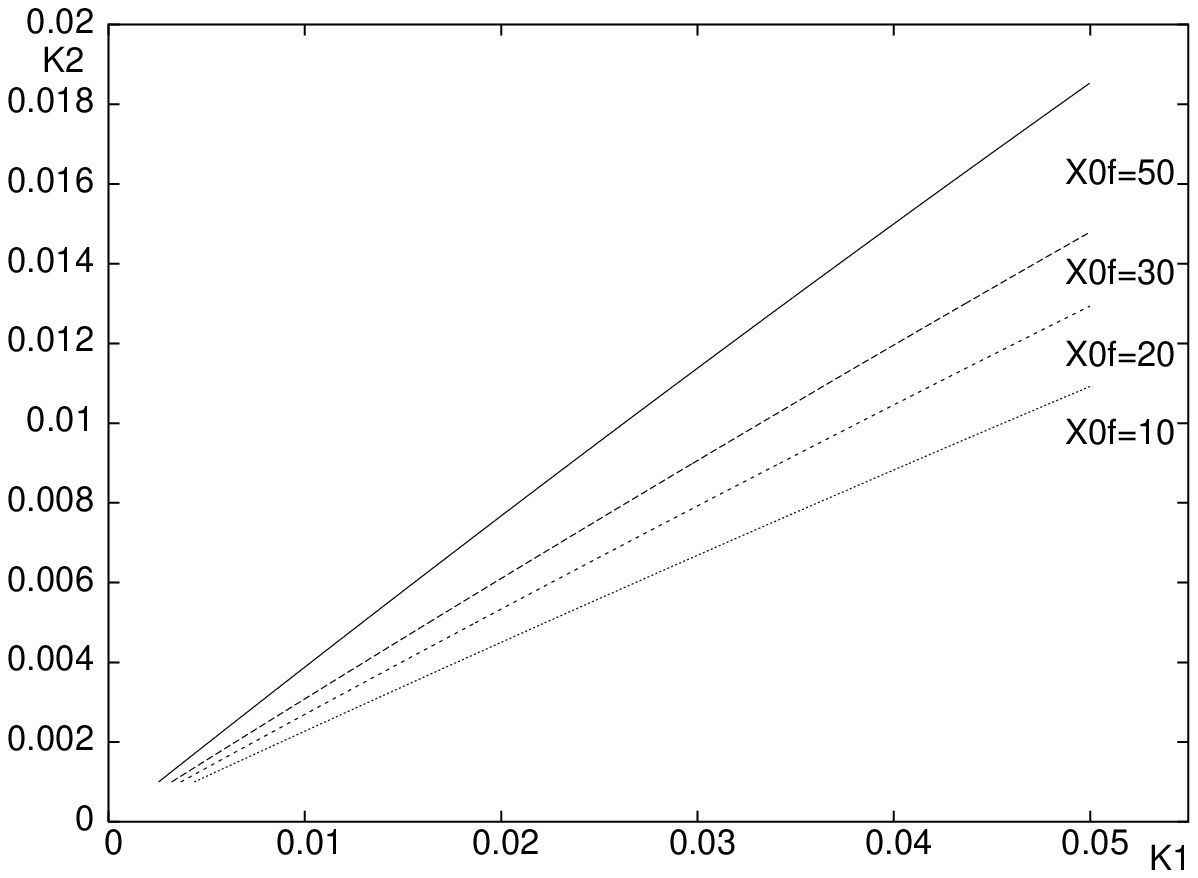}
\caption[]{${K2}$ against ${K1}$ for ${e=0.4}$ and for which ${A
-B=0}$.  The right graph is the same as the left one, but for 
${K1 \leq 0.05}$.}
\end{center}
\end{figure}

\begin{figure}
\begin{center}
\includegraphics[width=80mm,height=60mm]{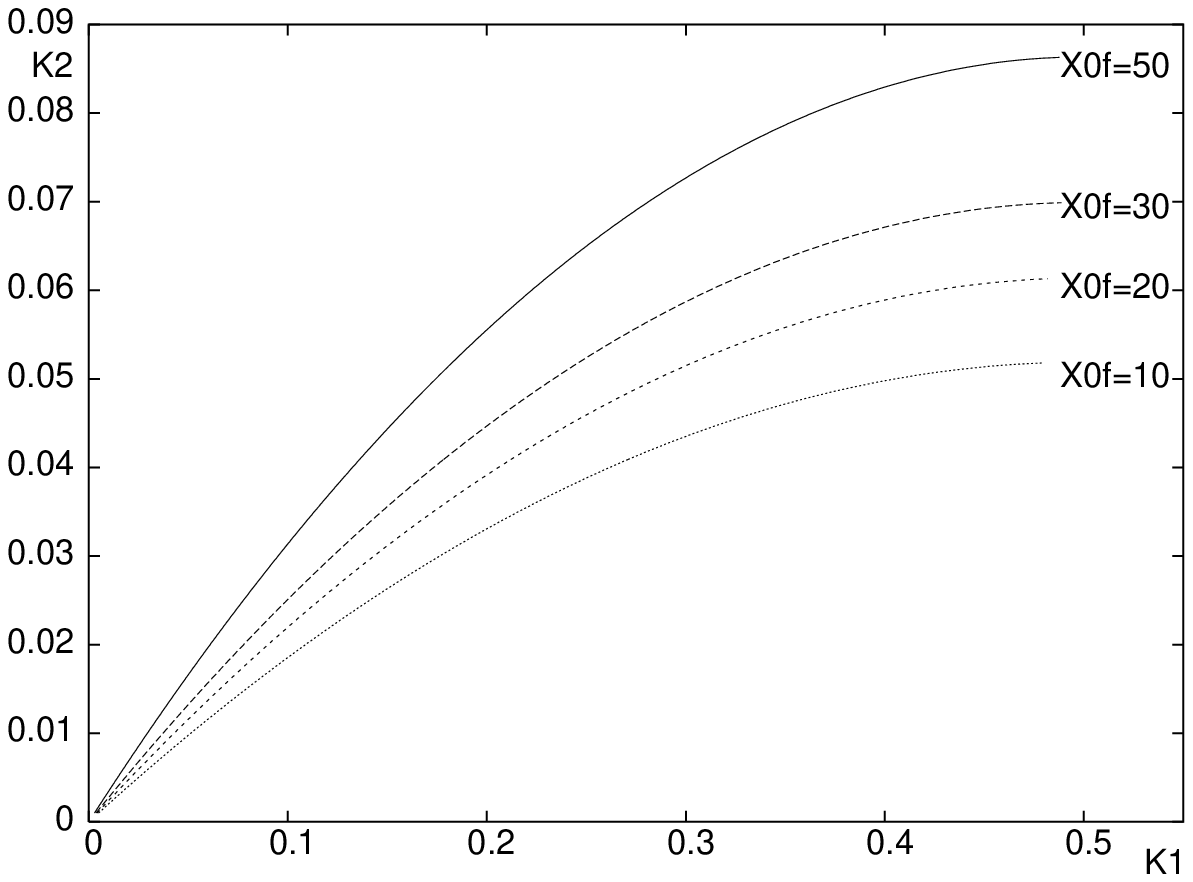}
\includegraphics[width=80mm,height=60mm]{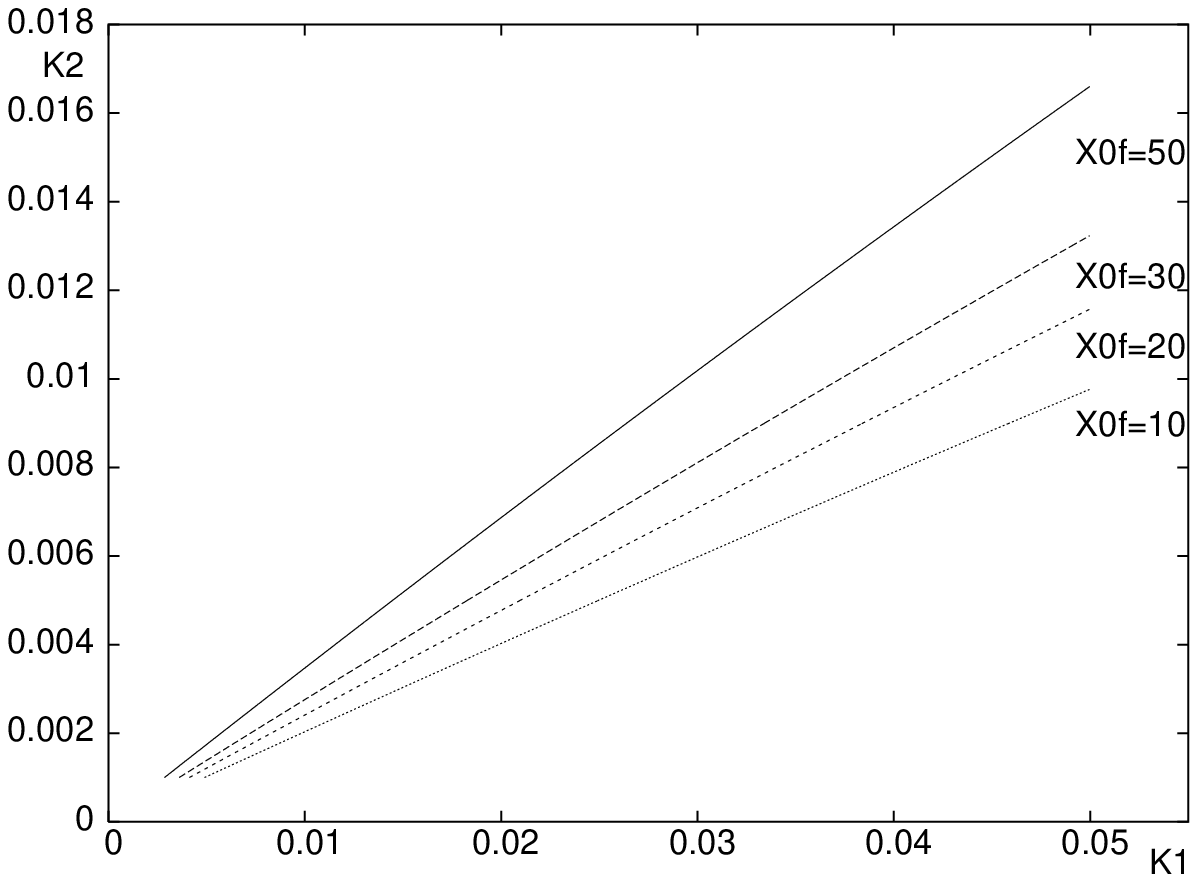}
\caption[]{${K2}$ against ${K1}$ for ${e=0.75}$ and for which ${A
-B=0}$.  The right graph is the same as the left one, but for 
${K1 \leq 0.05}$.}
\end{center}
\end{figure}

\subsection{SHORT PERIOD EVOLUTION} 

In order to be consistent with papers HTS1 and HTS2, and for 
completeness reasons, we also present results from comparing the
numerical and theoretical inner eccentricity, on shorter 
timescales. The results, presented in Tables 4,5 and 6, show the
percentage error between the averaged numerical and theoretical inner
eccentricity (equations (30) and (31) of HTS1 with the modification 
mentioned in section \ref{tef} and equations (28) and (29) of HTS2).  
Again, there are four entries per ${(K1-K2)}$ pair, corresponding, from 
top to bottom, to ${X=10,20,30,50}$ respectively.  The integrations 
were done with ${\phi=90^{\circ}}$ for a circular outer binary and with 
${f_{0}=90^{\circ}}$ and ${\varpi=0^{\circ}}$ for a system with an eccentric 
outer binary.  However, this does not affect the qualitative understanding
of the problem at all; similar results  are obtained for different initial
conditions and an example of that is given in Table 7.

\begin{table}
\caption[]{Percentage error between the
averaged numerical and averaged theoretical ${e_{1}}$.  The
theoretical model is based on equations (30) and
(31) of HTS1.  For all systems, ${e=0}$ and ${\phi=90^{\circ}}$.}
\vspace{0.1 cm}
\begin{center}	
{\normalsize \begin{tabular}{c c c c c c c }\hline
${K2\backslash\ K1}$ & ${0.001}$ & ${0.005}$ & ${0.01}$ & ${0.05}$ & ${0.1}$ & ${0.5}$ \\
\hline
0.001 & 4.9 & 4.7 & 5   & 5.1 & 5.3 & 5.2 \\
      & 1.6 & 1.6 & 1.6 & 1.6 & 1.7 & 1.4 \\
      & 0.9 & 0.9 & 0.9 & 0.9 & 0.9 & 0.6 \\
      & 0.4 & 0.4 & 0.4 & 0.4 & 0.4 & 0.3 \\
      
0.01 & 5.2 & 5.2 & 5.2 & 5.3 & 5.5 & 5.4 \\
     & 1.7 & 1.7 & 1.7 & 1.7 & 1.7 & 1.4 \\
     & 0.9 & 0.9 & 0.9 & 0.9 & 0.9 & 0.7 \\
     & 0.5 & 0.5 & 0.5 & 0.5 & 0.5 & 0.3 \\
   
0.1  & 7.1 & 7.1 & 7.1 & 7.2 & 7.3 & 7.1 \\
     & 2.6 & 2.6 & 2.6 & 2.6 & 2.6 & 2.2 \\
     & 1.6 & 1.6 & 1.6 & 1.5 & 1.5 & 1.1 \\
     & 0.9 & 0.9 & 0.9 & 0.8 & 0.8 & 0.5 \\
    
1    &17.6 &17.6 &17.6 &17.5 & 17.2 & 16 \\
     & 7.3 & 7.2 & 7.2 & 7.1 & 6.9 &  5.9 \\
     & 4.6 & 4.6 & 4.6 & 4.5 & 4.3 &  3.4 \\
     & 2.7 & 2.7 & 2.7 & 2.6 & 2.5 &  1.8 \\

10  &27.7 &27.7 &27.6 & 27.4 &27.2 &26.4  \\
    & 11  & 11  & 11  & 10.9 &10.7 & 9.9  \\
    & 6.8 & 6.7 & 6.7 & 6.6  & 6.5 & 5.8 \\
    & 3.8 & 3.8 & 3.8 & 3.7  & 3.6 & 3.1 \\
  
100  & 28.5 & 28.5 &28.5 &28.5 &28.4 & 28.4 \\
     & 11.1 & 11.1 &11.1 & 11  & 11  & 10.8 \\
     & 6.6  & 6.6  & 6.6 & 6.5 & 6.5 & 6.3 \\
     & 3.6  & 3.6  & 3.6 & 3.6 & 3.5 & 3.4 \\
   
1000 & 28.5 & 28.5 &28.5 &28.5 &28.5 & 28.6 \\
     & 10.9 & 10.9 &10.9 &10.9 &10.9 & 10.8 \\
     & 6.4  & 6.4  & 6.4 & 6.4 & 6.4 & 6.4 \\
     & 3.5  & 3.5  & 3.5 & 3.5 & 3.4 & 3.4 \\
     
\hline
\end{tabular}}
\end{center}	
\end{table}

\begin{table}
\caption[]{Percentage error between the
averaged numerical and averaged theoretical ${e_{1}}$.  The
theoretical model is based on equations (28) and
(29) of HTS2.  For all systems, ${e=0.4}$, ${f_{0}=90^{\circ}}$ and ${\varpi=0^{\circ}}$.}
\vspace{0.1 cm}
\begin{center}	
{\normalsize \begin{tabular}{c c c c c c c }\hline
${K2 \backslash\ K1}$ & ${0.001}$ & ${0.005}$ & ${0.01}$ & ${0.05}$ & ${0.1}$ & ${0.5}$ \\
\hline
0.001 & 2.5 & 2.5 & 2.5 & 2.5 & 2.5 & 2.2 \\
      & 0.8 & 0.8 & 0.8 & 0.8 & 0.7 & 0.6 \\
      & 0.4 & 0.4 & 0.4 & 0.4 & 0.4 & 0.3 \\
      & 0.2 & 0.2 & 0.2 & 0.2 & 0.2 & 0.1 \\
      
0.01 & 2.7 & 2.7 & 2.7 & 2.6 & 2.6 & 2.3 \\
     & 0.9 & 0.9 & 0.8 & 0.8 & 0.8 & 0.6 \\
     & 0.5 & 0.5 & 0.5 & 0.5 & 0.4 & 0.3 \\
     & 0.2 & 0.2 & 0.2 & 0.2 & 0.2 & 0.1 \\
   
0.1  & 3.8 & 3.7 & 3.7 & 3.7 & 3.6 & 2.9 \\
     & 1.5 & 1.5 & 1.5 & 1.4 & 1.4 & 1   \\
     & 1   & 1   & 1   & 0.9 & 0.9 & 0.5 \\
     & 0.6 & 0.6 & 0.6 & 0.5 & 0.5 & 0.2 \\
    
1    & 9.1 & 9.1 & 9   & 8.7 & 8.4 & 6.3 \\
     & 4.5 & 4.5 & 4.5 & 4.3 & 4   & 2.7 \\
     & 3.1 & 3.1 & 3   & 2.9 & 2.7 & 1.5 \\
     & 2   & 2   & 2   & 1.9 & 1.7 & 0.9 \\

10  & 12.8 & 12.8 & 12.7 & 12.5 & 12.1 & 10.4  \\
    & 5.9  & 5.9  & 5.9  & 5.7  & 5.5  & 4.5  \\
    & 3.8  & 3.8  & 3.8  & 3.7  & 3.5  & 2.6 \\
    & 2.4  & 2.4  & 2.4  & 2.3  & 2.2  & 1.5 \\
  
100  & 12.1 & 12.1 & 12.1 & 12  & 11.9 & 11.2 \\
     & 5.3  & 5.3  & 5.3  & 5.2 & 5.2  & 4.9 \\
     & 3.2  & 3.2  & 3.2  & 3.2 & 3.1  & 2.8 \\
     & 1.9  & 1.9  & 1.9  & 1.9 & 1.8  & 1.6 \\
   
1000 & 11.7 & 11.7 & 11.6 & 11.6 & 11.6 & 11.3 \\
     & 5    & 5    & 5    & 5    & 5    & 4.9 \\
     & 3    & 3    & 3    & 3    & 2.9  & 2.8 \\
     & 1.7  & 1.7  & 1.7  & 1.7  & 1.7  & 1.6 \\
\hline
\end{tabular}}
\end{center}	
\end{table}

\begin{table}
\caption[]{Percentage error between the
averaged numerical and averaged theoretical ${e_{1}}$.  The
theoretical model is based on equations (28) and
(29) of HTS2.  For all systems, ${e=0.75}$, ${f_{0}=90^{\circ}}$ and ${\varpi=0^{\circ}}$.}
\vspace{0.1 cm}
\begin{center}	
{\normalsize \begin{tabular}{c c c c c c c }\hline
${K2 \backslash\ K1}$ & ${0.001}$ & ${0.005}$ & ${0.01}$ & ${0.05}$ & ${0.1}$ & ${0.5}$ \\
\hline
0.001 & -0.1 & -0.1 & -0.1  &  0.2 & 0.6  & 0.5 \\
      & -0.7 & -0.6 & -0.6  & -0.4 & -0.2 & 0.1 \\
      & -1.2 & -1.2 & -1.2  & -0.9 & -0.6 & 0.1 \\
      & -2   & -1.9 & -1.9  & -1.6 & -1.2 & 0.1 \\
      
0.01 & -0.1 & -0.1 & 0.1  &  0.2 & 0.7  & 0.6 \\
     & -0.6 & -0.6 & -0.6 & -0.4 & -0.1 & 0.1 \\
     & -1.2 & -1.2 & -1.1 & -0.9 & -0.6 & 0.1 \\
     & -1.9 & -1.9 & -1.9 & -1.6 & -1.2 & 0.1 \\
   
0.1  & 0.4  & 0.4  & 0.4  & 0.7  & 1.1  & 1.1 \\
     & -0.3 & -0.3 & -0.3 & -0.1 & 0.1  & 0.4 \\
     & -0.9 & -0.9 & -0.9 & -0.6 & -0.4 & 0.2 \\
     & -1.7 & -1.7 & -1.6 & -1.3 & -1   & 0.1 \\
    
1    & 2.3  & 2.3  & 2.3  & 2.7  & 3.1  & 3.7 \\
     & 0.8  & 0.8  & 0.9  & 1    & 1.1  &  1.6 \\
     & 0.2  & 0.2  & 0.2  & 0.3  & 0.5  &  1   \\
     & -0.6 & -0.5 & -0.5 & -0.3 & -0.1 &  0.6  \\

10  & 4.6 & 4.7 & 4.7 & 4.9 & 5.1 & 6.4  \\
    & 1.9 & 1.9 & 2   & 2   & 2.1 & 2.8  \\
    & 1.2 & 1.2 & 1.2 & 1.2 & 1.3 & 1.8 \\
    & 0.5 & 0.5 & 0.5 & 0.6 & 0.6 & 1.1 \\
  
100  & 5.9 & 5.9 & 5.9 & 6    & 6.1 & 6.8 \\
     & 2.5 & 2.5 & 2.5 & 2.6  & 2.6 & 3.1 \\
     & 1.6 & 1.6 & 1.6 & 1.6  & 1.7 & 2    \\
     & 0.9 & 0.9 & 0.9 & 0.9  & 0.9 & 1.2 \\
   
1000 & 6.4 & 6.4 & 6.4 & 6.5 & 6.5 & 6.9 \\
     & 2.8 & 2.8 & 2.8 & 2.9 & 2.9 & 3.1 \\
     & 1.8 & 1.8 & 1.8 & 1.8 & 1.8  & 2  \\
     & 1   & 1   & 1   & 1   & 1.1  & 1.2 \\

\hline
\end{tabular}}
\end{center}	
\end{table}

\begin{table}
\caption[]{Percentage error between the
averaged numerical and averaged theoretical ${e_{1}}$, for a system
with ${K1=0.01}$, ${K2=1}$ and ${e=0}$.}
\vspace{0.1 cm}
\begin{center}	
{\normalsize \begin{tabular}{c c c c c}\hline
${X_{0} \backslash\ \phi}$ & ${0^{\circ}}$ & ${90^{\circ}}$ & ${180^{\circ}}$ & ${270^{\circ}}$ \\
\hline
10  & 7.2 & 17.6 & 5.3 & 17.7 \\
20  & 3.4 & 7.2  & 4.5 & 7.2 \\
30  & 2.3 & 4.6  & 4.4 & 4.6 \\
50  & 1.5 & 2.7  & 3.4 & 2.7 \\
\hline
\end{tabular}}
\end{center}	
\end{table}

\section{Applications in exosolar planetary systems: the HD217107 system}

In the past decade, 155  planets have been discovered orbiting
stars other than our Sun, with properties that are somehow
different compared to our solar system (e.g. eccentric orbits
are rather common among exoplanets).  Among them, multiple planetary
systems have been detected around 17 stars, by use of the Doppler 
technique.  For a summary of those developments up to date, see
Marcy et al. 2005.  Here, we discuss some possible applications of 
our formula on
exoplanets, although that was not our initial intention when we
started this work, firstly presented in HTS1.  

From the multiple exoplanetary systems that have been discovered so far, we 
picked up HD217107,
which has a dynamical setup similar to the one we study in
our paper (well
separated components, no mean motion commensurabilities).
So far, two planets have been detected orbiting HD217107 
(Fischer et al. 1999, 2001; Vogt et al. 2005).  The orbital elements of the system are given in Table 8.

Assuming that the two binaries move in coplanar orbits and substituting 
the HD217107 orbital elements into formula 
(\ref{final2}), we get  that ${\sqrt{\overline{e_{1}^{2}}}
\approx 0.025}$ (Keck) and ${\sqrt{\overline{e_{1}^{2}}}\approx 0.028}$ (Lick) 
(with ${\sin{i}=1}$, ${i}$ being the inclination of the orbital plane with the plane of the sky).  
Fig. 4 is a plot of inner eccentricity against time
for a system with the orbital parameters of HD217107 and it is obtained 
by integrating the full equations of motion numerically using our 
symplectic code (${e_{10}=0}$, ${f_{0}=0^{\circ}}$,
${\varpi=0^{\circ}}$).  As one can see, the maximum eccentricity is around
0.04, a value that is much smaller than the observed one
of around 0.13 (our theoretical model produces the same graphs).  
We would probably be able to have an orbit with a larger
eccentricity if we started the system with a small but non zero eccentricity,
e.g. 0.08 would give us a maximum eccentricity around 0.13.  However, this may
 not be a good assumption to make, as the planet is very close to the star 
(${a=0.074 AU}$) and tidal circularisation would be expected to take place.  Therefore, 
we would still need a not so small perturbation to the inner orbit, capable 
of maintaining an 
eccentricity of at least 0.13 (we say at least, because we do not really know 
which phase of the eccentricity oscillation we currently see).  We also performed integrations varying 
${\sin{i}}$, but with very little effect on the eccentricity. What seemed to affect things,
was to consider a system with a larger outer eccentricity (for example, for 
${e_{{\rm out}}=0.8}$, the maximum ${e_{{\rm in}}}$ was around 0.12).

Another possibility of course, is that the two orbits are not coplanar.  
In the case of low mutual inclination ${I}$ (${I<39.23^{\circ}}$ or 
${I>140.77^{\circ}}$), our past experience from circular binaries 
(Georgakarakos 2004) and some quick numerical integrations of the
secular equations of motion with a 4th-order Runge-Kutta method with
variable stepsize (Press et al. 1996; for the secular equations see
Marchal 1990), led us to the conclusion that there was little difference 
between the eccentricity of a coplanar and a non coplanar orbit, with the
rest of the orbital elements being the same (an exception here might be
the occurence of a secular resonance between the two pericentre frequencies, which
can increase the amplitude of the eccentricity oscillation).  
On the other hand, in
the case of high mutual inclination (${39.23^{\circ}<I<140.77^{\circ}}$), 
it is known that the eccentricity can reach significant values, it
can even become one for ${I=90^{\circ}}$ (${e_{{\rm max}}=\sqrt{1-\frac{5}{3}
\cos^{2}{I}}}$; more about the high and low inclination regimes in 
Kozai 1962).  

Thus, as a conclusion, we could say that with the current information about
the HD217107 system, the planets that have been detected there so far is not
very likely to move on the same plane.  Unfortunately, the two formulae presented
in this paper (eqns. (\ref{final1}) and (\ref{final2})) are only applicable for 
hierarchical triple systems on coplanar orbits (or orbits with a mutual inclination
of a few degrees) and hence, they can not be used to place further constraints to the 
system.

We would like to point out here that equations (1) and (2) are purely classical and do not take into account any relativistic effects.  However, in  situations where a planet is very close to the host star, relativistic apsidal precession of the inner orbit could affect the evolution of eccentricity (e.g. see Holman et al. 1997, Ford et al. 2000, Mardling ${\&}$ Lin 2002).

\begin{table}
\caption[]{Orbital elements for the planetary system HD217107, taken by Vogt et al. 2005. 
The periods are in days, the planetary masses in Jupiter 
masses and the the semimajor axes in AU.  Keep in mind that the planet masses given in the table are not
the actual masses, but ${M_{planet}\sin{i}}$, where the angle ${i}$ is the 
inclination  of the orbital plane with the plane of the sky.} 
\vspace{0.1 cm}
\begin{center}	
{\normalsize \begin{tabular}{c c c c c}\hline
 Object & Mass & Period & Semimajor axis & Eccentricity \\
\hline
HD217107         & ${1.053M_{\odot}}$ & --- & --- & ---\\
HD217107b        & ${1.37M_{J}}$      & 7.1269d & 0.074 & 0.13 \\
HD217107c (Keck) & ${2.1M_{J}}$       & 3150d & 4.3 & 0.55 \\
HD217107c (Lick) & ${3.31M_{J}}$      & 2465d & 3.6 & 0.53 \\
\hline
\end{tabular}}
\end{center}	
\end{table}

\begin{figure}
\begin{center}
\includegraphics[width=80mm,height=60mm]{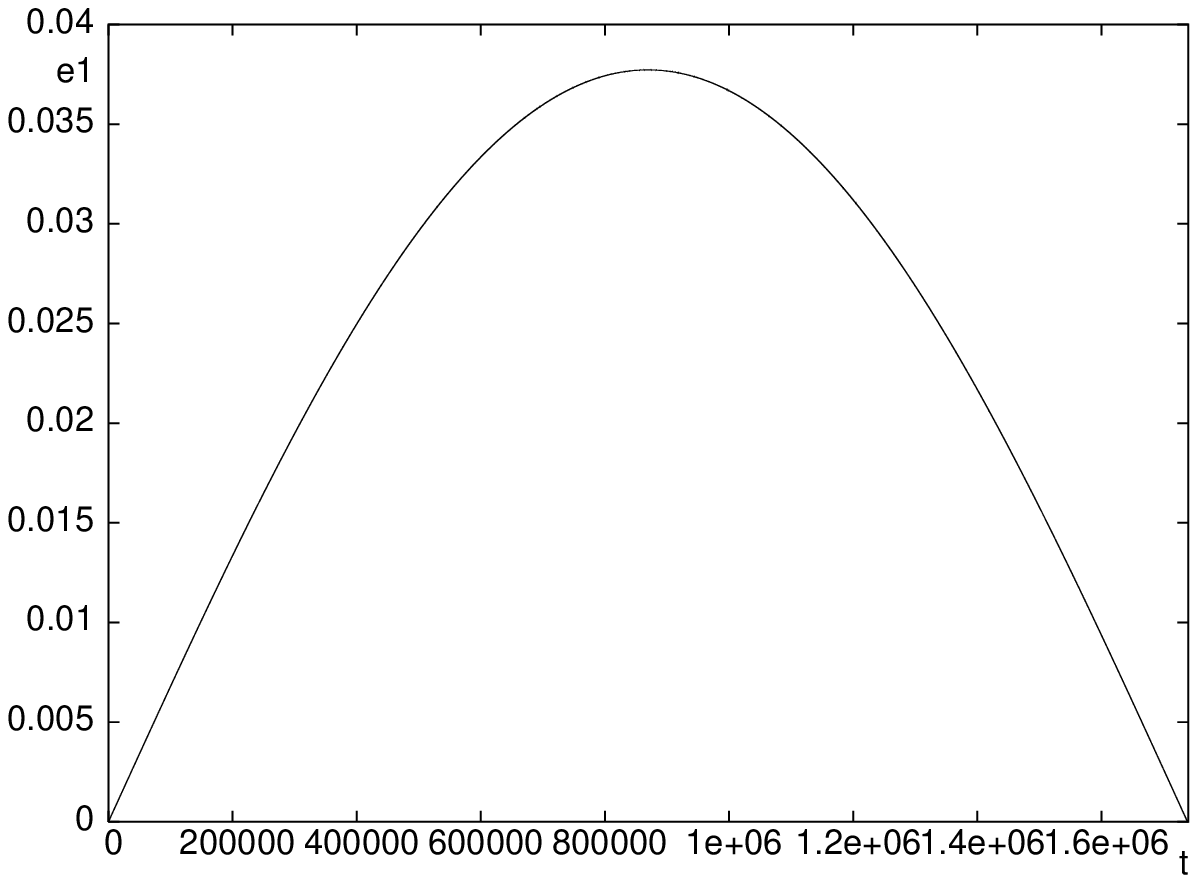}
\includegraphics[width=80mm,height=60mm]{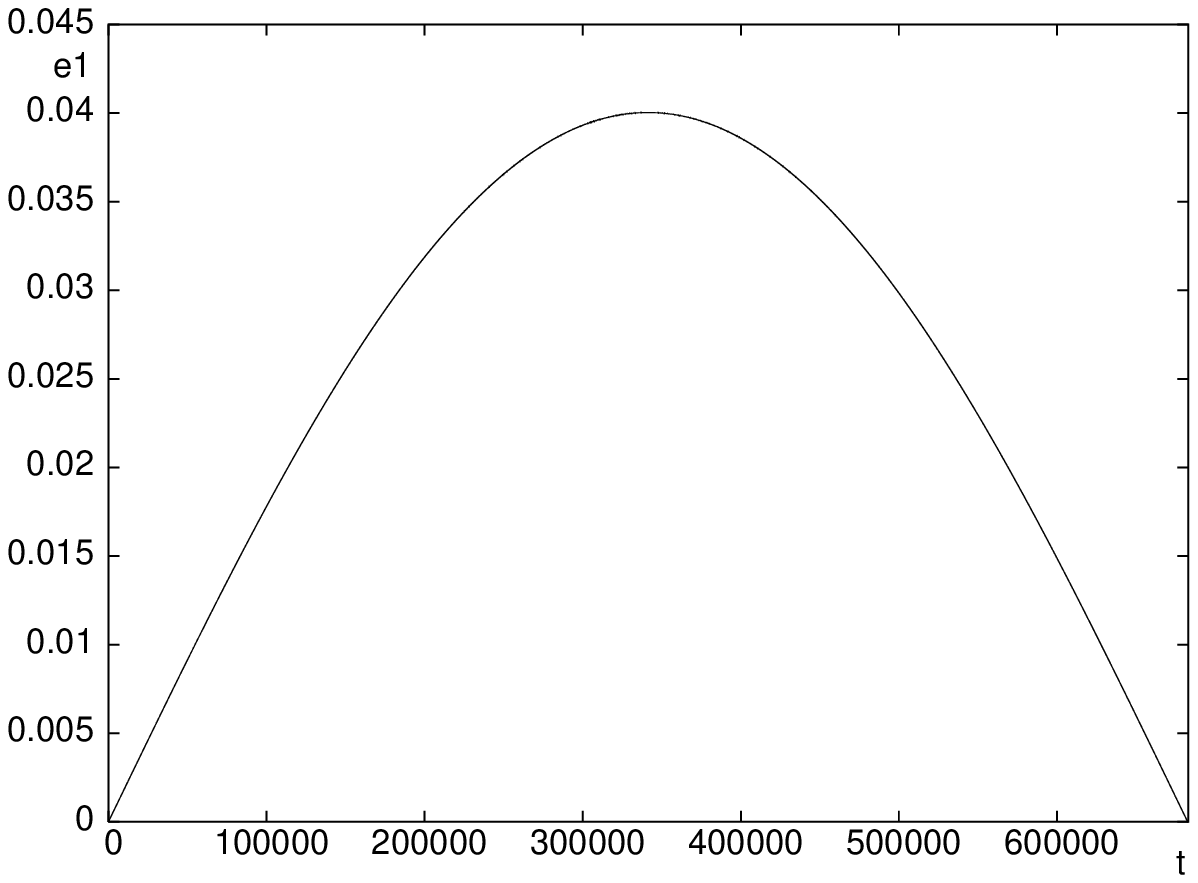}
\caption[]{Inner eccentricity against time for HD217107.  The graphs
come from the numerical integration of the full equations of motion (the left graph is based on the Keck Observatory data, while the left one uses data from Lick Observatory).   The
initial conditions are ${e_{10}=0}$, ${f_{0}=0^{\circ}}$ and  ${\varpi=0^{\circ}}$.  The time ${t}$ is in yrs (the period of oscillation for the left graph is ${1.74 10^{6}}$ yrs and ${6.83 10^{5}}$ yrs for the right one).}
\end{center}
\end{figure}

\section{Summary}

In two previous papers, we derived formulae for estimating the inner 
eccentricity in hierarchical triple systems with coplanar orbits and with 
the inner eccentricity being initially zero.  However, those calculations 
were done for systems with comparable masses.  In the present paper, 
we tested the formulae for systems with mass ratios from 
${10^{-3} \hspace{0.2cm}\mbox{to} \hspace{0.2cm}10^{3}}$.   
The theoretical models appeared to work 
well for most of the cases.  There were a few cases for which the theory
did not work well, but no theory is perfect.  Finally, we applied 
the theory on the exosolar planetary system HD217107, in an attempt
to obtain more information about its orbital characteristics, although
that kind of application was not part of the  initial motivation for the 
construction of the models.  We concluded that the two planets orbiting the host star are probably on non coplanar orbits.  The derivation of a three dimensional formula
for eccentric binaries is one of our future aims, as it would definitely 
prove more helpful for various dynamical problems, especially with the
constant discovery of more and more exosolar planets every day.

\section*{ACKNOWLEDGMENTS}

The author wants to thank Douglas Heggie for the useful
discussion concerning this paper and Seppo
Mikkola, who kindly provided the code for integrating hierarchical
triple systems.

\section*{REFERENCES}
 
Fekel F. C., Jr., Tomkin J., 1982, ApJ 263, 289\\
Fischer D. A., Marcy G. W., Butler R. P., Vogt S. S., Apps K., 1999, 
PASP, 111, 50\\
Fischer D. A., Marcy G. W., Butler R. P., Vogt S. S., Frink S., Apps K., 2001, 
ApJ, 551, 1107\\
Ford E. B., Kozinsky B., Rasio F. A., 2000, ApJ, 535, 385\\
Georgakarakos N., 2002, MNRAS, 337, 559\\
Georgakarakos N., 2003, MNRAS, 345, 340\\
Georgakarakos N., 2004, CeMDA, 89, 63\\
Georgakarakos N., 2005, MNRAS, 362, 748\\
Hinkle K. H., Fekel F. C., Johnson D. S., Scharlach W. W. G., 1993, AJ, 105, 1074\\Jha S., Torres G., Stefanik R. P., Latham D. W., Mazeh T., 2000, MNRAS, 317, 375\\
Holman M., Touma J., Tremaine S.,1997, Nat, 386, 254\\
Kozai Y., 1962, AJ, 67, 591\\
Marchal C., 1990, The Three-Body Problem.  Elsevier Science Publishers, the Netherlands\\
Marcy G. W., Butler R. P., Fischer D. A., Vogt S. S., Wright J. T., Tinney C. G., Jones H. R. A., 2005, PThPS, 158, 24\\ 
Mardling R.A., Lin D.N.C., 2002, ApJ, 573, 829\\ 
Mikkola S., 1997, CeMDA, 67, 145\\
Press W. H., Teukolsky S. A., Vetterling W. T.,
Flannery B. P., 1996, Numerical Recipes In Fortran 77 (2nd
ed.).  Cambridge Univ. Press, NY\\
Vogt S. S.,  Butler R. P, Marcy G. W., Fischer D. A., Henry G. W., Laughlin G., Wright J. T., Johnson J. A., 2005, ApJ, 632, 638
\end{document}